\documentclass[12pt]{article}
\usepackage{stix2}
\usepackage[utf8]{inputenc}
\usepackage{graphicx} 
\usepackage{amsmath, amssymb} 
\usepackage{natbib} 
\usepackage{setspace}
\usepackage{doi}

\usepackage{geometry}

\title{GPC/m: Global Precipitation Climatology by Machine Learning; Quasi-global, Daily, and One Degree Spatial Resolution}
\author{Hiroshi G. Takahashi \\
        \small Tokyo Metropolitan University \\
        \small \texttt{hiroshi3@tmu.ac.jp}}
\date{\today}

\begin{document}
\maketitle

\begin{abstract}
This paper presents a new precipitation dataset that is daily, has a spatial resolution of one degree on a quasi-global scale, and spans more than 42 years, using machine learning techniques. The ultimate goal of this dataset is to provide a homogeneous daily precipitation dataset for several decades without gaps, which is suitable for climate analysis. As a first step, 42 years of daily precipitation data was generated using machine learning techniques. The machine learning methods are supervised learning, and the reference data are estimated precipitation datasets from 2001 to 2020. Three machine learning methods are used: random forest, gradient-boosted decision trees, and convolutional neural networks. The input data are satellite observations and atmospheric circulations from reanalysis, which are somewhat modified based on knowledge of the climatological background. Using the trained statistical models, we predict back to 1979, when daily precipitation data was almost unavailable globally. The detailed procedures are described in this paper. The produced data have been partially evaluated. However, additional evaluations from different perspectives are needed. The advantages and disadvantages of this precipitation dataset are also discussed. Currently, this GPC/m precipitation dataset version is GPC/m-v1-2024.
\end{abstract}

\section{Introduction}
The importance of precipitation datasets has increased in recent years as they are extensively used for climate and meteorological analyses, such as precipitation variability, water budget, evaluation of climate model, and related research \citep[e.g.,][]{tre11}. However, the available precipitation data have some limitations, especially difficulties in using precipitation data globally, including over the ocean over long periods \citep{bec13}. This study produces a prototype of the precipitation dataset using machine learning methods. This machine learning-based precipitation can be used in climate and related studies. Additionally, this study can contribute to discussing not only the advantages but also the problems of global precipitation climatological datasets using machine learning methods. In general, these advantages are easy to understand, but this dataset also focuses more on limitations and issues with the use of machine-learning-based precipitation data as climate data \citep[e.g.,][]{kum15,tak23}.

Global Precipitation Climatology Centre (GPCC) and Global Precipitati,on Climatology Project (GPCP) are often used as de facto standard precipitation datasets with a focus on long-term climate variability. These datasets provide monthly moderate-resolution outputs that are sufficient for analyses on global and continental scales \citep{sch14}. The typical scales are monthly and with a 1$^{\circ}$ spatial resolution. Land datasets are available starting from the early 20th century although the spatial coverage and density of the observations are limited. High-resolution but period-limited datasets derived from rain-gauge observations became available starting in approximately 1950. Global datasets, including sea and no-ground observation areas, generally began in 1979, stemming from advances in satellite observations \citep[e.g.,][]{adl03}. Currently, the two dataset have been updated positively.

Recent datasets such as IMERG and GSMAP focus on severe meteorological disasters and have very high temporal and spatial resolutions. Typical scales are shorter than one hour and have 0.1$^{\circ}$ order resolution \citep[e.g.,][]{huf15}. Although precipitation data are widely available, these datasets are mainly from 2000 onward, with limited data available before 1998. In addition, station-based precipitation data are available but are limited to precipitation over land. Thus, precipitation data over oceans are lacking. For climate variation analysis using daily convective information, the outgoing longwave radiation (OLR) dataset has been used as the index of convective activity for more than 30 years \citep{lie96}. Although the OLR is not a problem, precipitation datasets are more likely to be required, particularly for water cycle analysis. OLR is regularly updated. However, updating the OLR dataset may not be so easy, even though OLR data are very popular. Similarly, maintaining the quality even with the same source of datasets from meteorological stations and satellites is difficult.

In the future, precipitation values in reanalysis datasets such as ERA5 \citep{her20} may come to be used as precipitation data in some fields. However, these data are only predicted by global atmospheric models and must be compared with observation values. Nevertheless, because the recent improvement in the quality of the reanalysis datasets is dramatic, the current situation is quite different from 10 years ago \citep{dee11}.

One of the main challenges of the climatological precipitation dataset is its long-term homogeneous quality. We want to extend the precipitation data back to 1979. However, gaps in the time axis need to be minimized. In the station-based data, the gaps may come from the changes in the local meteorological stations of precipitation. For satellite-derived datasets, the biases from inter-satellite gaps, even with the same or similar instruments, changes in satellite observations instrument itself, and so on. Reanalysis data included all the gaps in the observations, which were used as assimilations. Thus, reanalysis itself also has gaps in time directions.

Although it is difficult to completely remove gaps in precipitation datasets, they can be reduced using machine learning methods \citep[e.g.,][]{goo16} and climatological knowledge. Climatological homogeneity is insufficient at present and will be a challenge in the future; however, it can be mitigated using machine learning.

One of the major purposes of this study was to create a climatologically consistent dataset. We hope that the precipitation data will be easy to use for analysis. When the resolution of the target (reference) data and its features (input data) is high, a high-resolution dataset can be produced. The recently precipitation datasets are useful but their resolution is too high in some cases, although they can be used by averaging or smoothing.

In the algorithm, all datasets use statistical models, which may be well-designed based on empirical knowledge. However, if input data as features include important information, the machine learning approach can possibly provide datasets of similar quality to the target data from data-driven processes. Therefore, we attempted to develop a data-driven algorithm. This may be advantageous as machine learning can use information that scientists have not focused on and may avoid needless information that scientists take for granted but have not proven.

Our machine learning-based daily precipitation data with moderate resolution was named as global precipitation climatology by machine learning, "GPC/m".

\section{Methods and data} 
Machine learning methods can offer several advantages for data analysis in atmospheric and climate sciences. In particular, they provide a robust method for capturing the nonlinear dependencies and interactions among features. Because precipitation processes are highly nonlinear, machine learning methods can provide useful datasets and multilateral knowledge of precipitation processes and their estimations. Thus, this study attempted to produce a precipitation dataset using machine learning.

GPC/m was produced from OLR and atmospheric reanalysis datasets, such as horizontal winds, geopotential height, and water vapor, using machine learning-based statistical models. In this version of GPC/m, daily precipitation dates back to 1979. The decision to go back to 1979 was made because OLR has been mainly available since 1979, and satellite observations have been widely used for atmospheric reanalysis since 1979.

Training and predictions were conducted for individual horizontal grids, which means that all the individual grids had different statistical models. Another option is to use one large unified statistical model applicable for all grids that represents the relationships for all grids. This unified model strategy is similar to highly sophisticated dynamic models (e.g., global climate models). Constructing multiple regionally unified statistical models for similar climate conditions would also be possible. Thus, extremely various designs of statistical models are possible, which means that we have no clear answer, even in the near future. We selected individual statistical models for individual grids because one large unified statistical model requires huge computational costs for massive training, but we can use only limited computational resources. We should also focus on large statistical models that have been developing worldwide in the future.

Specifically, three machine learning methods were used for GPC/m: Random Forest (RF), eXtreme Gradient Boosting (XGB), and Convolutional Neural Networks (CNN). These three methods independently predicted precipitation values. GPC/m precipitation was computed using the simple ensemble mean of all three methods, which cancels out some of the climatological biases among the three predictions. In future studies, we will increase the number of ensembles, e.g., by changing the hyperparameters, which can be a type of parameter sweeping.

For training and predictions of the machine learning, we prepare the input datasets. The GSMaP satellite precipitation data \citep{kub07} were used as the target precipitation dataset, National Oceanic and Atmospheric Administration (NOAA) OLR as convective activity information, and atmospheric features from atmospheric reanalysis data \citep[JRA55][]{kob15}. All data were prepared on a daily basis and at 1$^{\circ}$ $\times$ 1$^{\circ}$ resolution. Basically, the input data and target grids are not fully matched and the grid sizes are different. The closest grid was assigned as the input data. The preliminary test showed that differences in the grid position and size had little effect on the results. For comparison, another major precipitation dataset, Integrated Multi-satellitE Retrievals for GPM \citep[IMERG;][]{huf15b}, was also used. 

The area of the GPC/m precipitation dataset is defined as zonal from 0.5$^{\circ}$E to 359.5$^{\circ}$E, that is, global, and meridional from 40.5$^{\circ}$S to 49.5$^{\circ}$N, that is, 360 $\times$ 91 grid points. The target area is also due to the training dataset of the GSMaP. Also, the precipitation estimation is basically related to the OLR (infrared radiation) input data, which is a better convective indicator over the tropical region. Because the statistical models were constructed for the individual grids, other input data, such as the geopotential height at 500 hPa, may be a better indicator of precipitation in the mid-latitudes; however, it is quite difficult to determine appropriate features for individual grid. We currently use the same features for all grid points.

The four steps in the basic procedure of the three machine learning methods for the GPC/m precipitation dataset are described below. In practice, however, the four steps, evaluation of the predictions of the models, and adjustment of the hyperparameters of the models were iterated before we determined the setting of the final provided dataset, GPC/m.

First, we prepared data for training, the input and reference precipitation data from 2001 to 2020, to develop a statistical model for individual grids. The input data were also prepared for a prediction period of 22 years from 1979 to 2000. Over these 22 years, GSMaP and other commonly used daily precipitation datasets do not provide daily precipitation values, particularly over the ocean.

Second, we performed training to construct statistical models for individual grids using the target and feature values daily for 20 years, from 2001 to 2020. All the statistical models were constructed for individual grids. Thus, the statistical models were independent of the grid. 

Third, we then predicted the precipitation values from 1979 to 2020 using input data without reference data and the trained statistical models for individual grids. Note that we also predicted from 2001 to 2020 using the trained statistical models for individual grids. Thus, the GPC/m values are predicted values, not reference values. Also, using a statistical model with input data, we predicted the precipitation values after 2020 if input data are available. For the test, we made predictions until 2023 using the input data and the trained models. The first through the third processes are performed using three different machine learning methods, respectively.

Fourth, the values predicted by the three methods were averaged and formatted as products. The three machine-learning methods are explained in detail below.

\subsection{Random Forest (RF)} 

RF uses major machine learning techniques that are also applied to atmospheric and climate sciences. \cite{bre01} provided detailed theoretical insights into the algorithm. \cite{hir19} estimated precipitation at high temporal resolution from geostationary satellite data using RF. Furthermore, \cite{lag17} applied the RF to predict severe weather events.

\subsubsection{Data Preparation and Preprocessing} 

The precipitation data were scaled to a 24-hour accumulation period and capped at 50 mm to eliminate rare extreme values, outliers, and errors. Atmospheric features included OLR, zonal and meridional wind components at 200 hPa, precipitable water, water vapor fluxes, stream function at 850 hPa, surface air temperature, and geopotential height at 500 hPa.

This study used a climatological analysis knowledge. The absolute value of OLR is generally not useful for understanding of convective conditions because the OLR data are provided mainly at the level of the top cloud, which can be various in seasonal march. To include weather time-scale information, we applied the temporal high-pass filter (25-day) for the timeseries of OLR at each grid, subtracting a 25-day running mean. Anomalous OLRs and other meteorological values are commonly used in the analysis of climatology and meteorology. This preprocessing and the choice of meteorological variables greatly improved the precipitation prediction based on the vast amount of preliminary analysis.

\subsubsection{Configuration and Training} 
The RF regression model was based on the RandomForestRegressor class of the sklearn.ensemble library (Pedregosa et al., 2011). The model parameters were configured to optimize the accuracy and complexity and prevent overfitting. We set \textit{n\_estimators} to 200, which enhances model stability and accuracy by determining the number of trees constructed during the boosting process. A parameter \textit{max\_depth} was established at 10 to control the complexity of the model and avoid overfitting by limiting the depth of each tree. Parameters \textit{min\_samples\_split} and \textit{min\_samples\_leaf} were set to 10 and 5, respectively, ensuring sufficient sample sizes in the nodes and leaves, which helps in preventing the creation of overly complex trees. Additionally, \textit{bootstrap} was enabled, further assisting in generalization. 

In the product run, training was performed from January 2001 to December 2020. Prior to the product run, we had divided the 20-year data into training data and evaluation data, reorganized the model, and improved the model's performance. For example, we trained the statistical model using 2001--2016 data and evaluated the model's performance by comparing it with observations of 2017--2020 period. These development processes were iteratively performed to improve model performance.

The model performance was also evaluated in terms of the spatial patterns of precipitation and the daily precipitation sequence by low-pressure systems, as presented in Section 3. However, some aspects, such as long-term changes in quasi-global mean precipitation, have yet to be well represented in this version.

\subsubsection{Predictions}  
After training, the RF statistical models for the individual grids were preserved for later predictions during the prediction period. In the earlier stage, the predictions during the training period were assessed by comparing them with target data. Using the RF statistical models for the individual grids trained with 20-year data from 2001 to 2020, we predicted precipitation values from 1979 to 2020 and also preliminarily extended to 2022.

\subsection{eXtreme Gradient Boosting (XGB)}  
XGB is a gradient-boosted decision trees method, well-known for its speed and high performance. These methods also offer several advantages in capturing non-linear dependencies and interactions among features, which are significant in precipitation processes. XGB also includes advanced features that mitigate overfitting.

XGB has been applied for precipitation prediction in previous studies. \cite{don23} used XGB to predict precipitation from downscaled and bias-corrected numerical weather predictions. Also, \cite{lin23} used machine learning models, including RF and XGB, to predict and analyze extreme precipitation intensity and frequency in various regions of the United States.

\subsubsection{Data Preparation and Preprocessing}  
The input data for XGB were almost the same as those of the RF. However, the surface air temperature, which was determined by trial and error, was not used in XGB.

\subsubsection{Configuration and Training}  
XGB was meticulously configured with the following hyperparameters to optimize the performance and minimize overfitting. We set the \textit{n\_estimators} to 256 to define the number of boosting rounds or trees, which is a critical factor for capturing complex patterns and enhancing model stability. The \textit{learning\_rate} was set to 0.16, controlling the step size at each iteration and playing a vital role in reducing overfitting while ensuring better convergence. We limited the depth of each tree to a \textit{max\_depth} of 16, allowing the model to capture detailed data patterns while maintaining balance to prevent overfitting. The \textit{gamma} parameter was set to 20, which specifies the minimum loss reduction required for further partitioning on a leaf node, thus helping control overfitting by restricting the algorithm. Similar to RF, development processes were repeated to improve model performance.

\subsubsection{Predictions}  
Similarly, after training, the predictions during the training period were assessed by comparing them with target data. Using the trained XGB statistical models, we predicted precipitation values from 1979 to 2020 and also preliminarily extended to 2022.

\subsection{Convolutional Neural Networks (CNN)}  
As the third method, this study also adopted a deep learning model-based approach, CNN. This method can also learn nonlinear relationships \citep[e.g.,][]{lec15}, which can be useful for meteorology, such as the relationship between precipitation and associated meteorological variables. Climate research also uses this CNN methodology \citep[e.g.,][]{liu16}. Because this method is quite different from the other two methods, it may provide a different prediction from the two methods, which can reduce the mean biases. 

\subsubsection{Data Preparation}  
The input data as target data for CNN were the same as RF and XGB. However, the meteorological variables of the input data as features, as six distinct channels, were basically the same as the RF except for not using the surface air temperature and geopotential height at 500 hPa. Also, the forms of the input feature data are different.

In CNN, we used two-dimensional data arrays similar to images. We prepared a matrix of 10 $\times$ 10 grid values centered on the grid point closest to the GPC/m product for the input data. Although the spatial resolution differs between the OLR and reanalysis data, resulting in varying spatial extents covered by the 10 $\times$ 10 grids, we opted not to adjust the grid positions. This decision was strategic, aiming to capture various meteorological information across multiple spatial scales. Also, the input data were normalized by the respective standard deviations. This approach enhances our ability to predict precipitation by integrating various scales and types of atmospheric data, thereby improving the accuracy and effectiveness of our predictive models in meteorology and climatology. Trial and error determined these detailed procedures because there was no simple logical support.

Same as RF and XGB, the OLR data were high-pass filtered OLR in CNN. This preprocessing and the choice of the meteorological variables are highly improved the precipitation prediction based on the vast amount of preliminary analysis.

\subsubsection{Model Construction} 
The sequential CNN model incorporates three convolutional layers, each followed by dropout layers, to mitigate overfitting, which is a common challenge in deep learning models, as highlighted by \cite{kri12} and \cite{sri14}. Also, specifically, we employed a linear activation function in the output layer for continuous outcome prediction, such as precipitation amounts, as a regression-oriented approach \citep{goo16}.

\subsubsection{Training and Evaluation}  
The CNN models were trained over 30 epochs with a batch size of 32. These procedures were determined to balance computational efficiency with learning effectiveness. Also, the following detailed designs were determined by considering trial-and-error and statistical characteristics of daily precipitation.

We employed a hybrid loss function as an evaluation metric for the CNN. This function is designed to apply asymmetrical penalties for prediction errors, with particular emphasis on the severity of penalties based on whether errors are positive or negative to consider the characteristics of daily precipitation.

For positive errors, a square root-transformed discrepancy was used for the evaluation. Conversely, the function directly applied penalties to negative errors based on the absolute differences. This method treats relatively heavier penalties for negative errors than for positive ones. Daily precipitation has many zero values, and the negative errors are more serious. This asymmetrical penalization may induce bias in the model, encouraging it to avoid underestimations rather than overestimation. To avoid such adverse effects, we carefully adjusted the model structure and hyperparameters while modifying the penalty.

\subsubsection{Prediction}  
Similarly, using the trained CNN statistical models, we predicted the daily precipitation from 1979 to 2020 and preliminarily extended it to 2022. Compared with the RF and XGB, CNN models have stronger underestimation biases. The negative bias in the quasi-global mean precipitation (Section 3.1) can be influenced by the negative biases of the CNN predictions from 1979 to 2000. Although we could not understand the cause of the negative bias during the prediction period, we reduced it as much as possible.

\section{Reproducibility of precipitation dataset} 

This section demonstarates reproducibility of the produced precipitation dataset "GPC/m", in terms of climatology, interannual variability, and significant past variations in precipitation. While we invite scientists, with the exception of the author of this paper, to further investigate various climatological phenomena from different perspectives than the author, we primarily evaluate some specific climatological results in previous papers by the author.

\subsection{Climatology} 

As the most basic reproducibility of the preciptiation dataset, the 22-year climatology of annual precipitation during the prediction period without observational target values shows a very similar spatial pattern in precipitation to the well-used GPCP climatology of the same period (Figure 1). Major spatial peaks are observed over the intertropical convergence zone (ITCZ) along the equator, the Asian monsoon regions, the Maritime continent, major tropical storm regions, and storm track regions in the mid-latitudes.
\begin{figure*}
      \centering
      \noindent\includegraphics[width=26pc,angle=270]{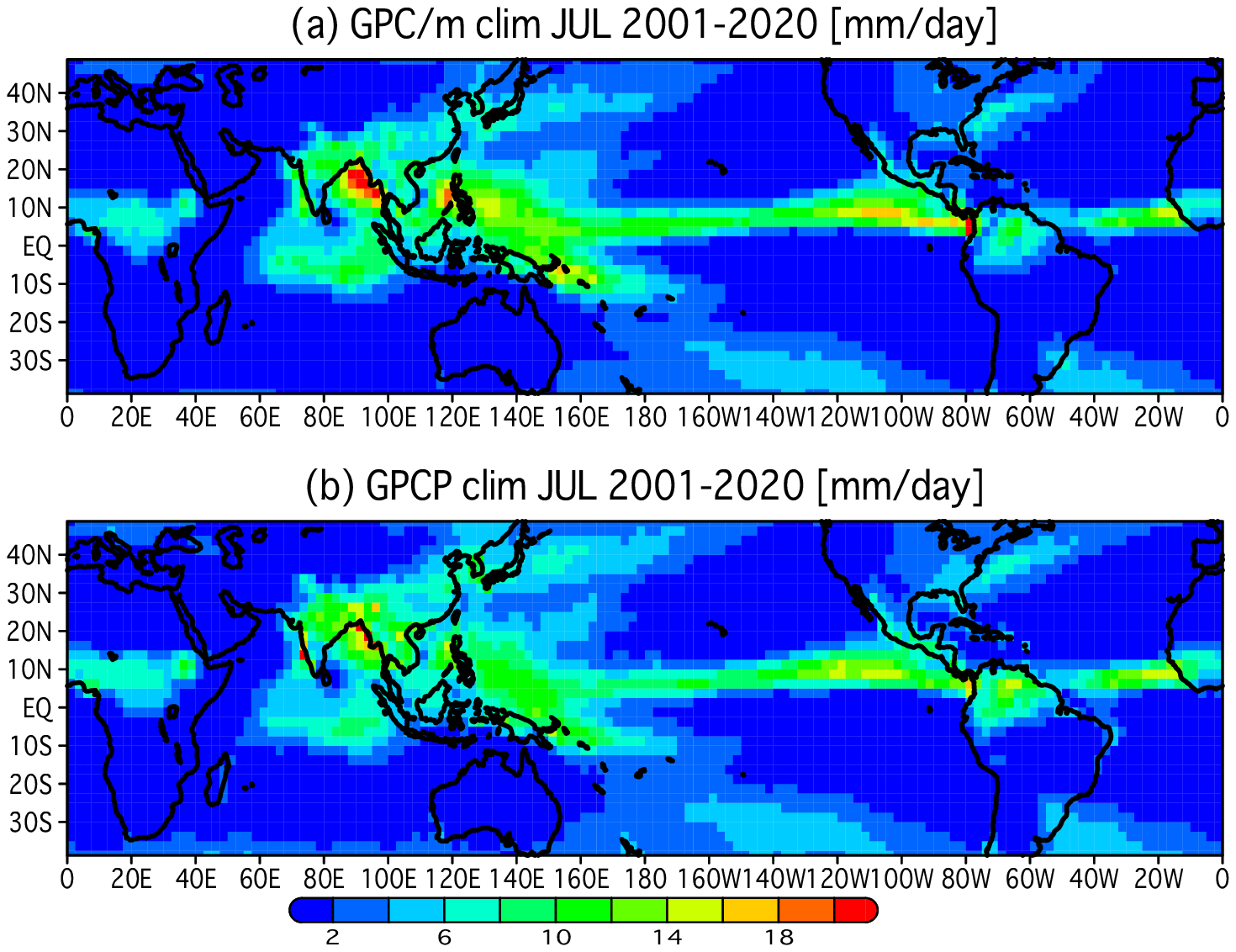}
      \caption{(a) Map of the climatological annual precipitation during the period from 2001 to 2020 in GPC/m. (b) Same as (a), but for GPCP. Units are mm day$^{-1}$.}
\end{figure*}

\begin{figure*}
      \centering
      \noindent\includegraphics[width=22pc,angle=270]{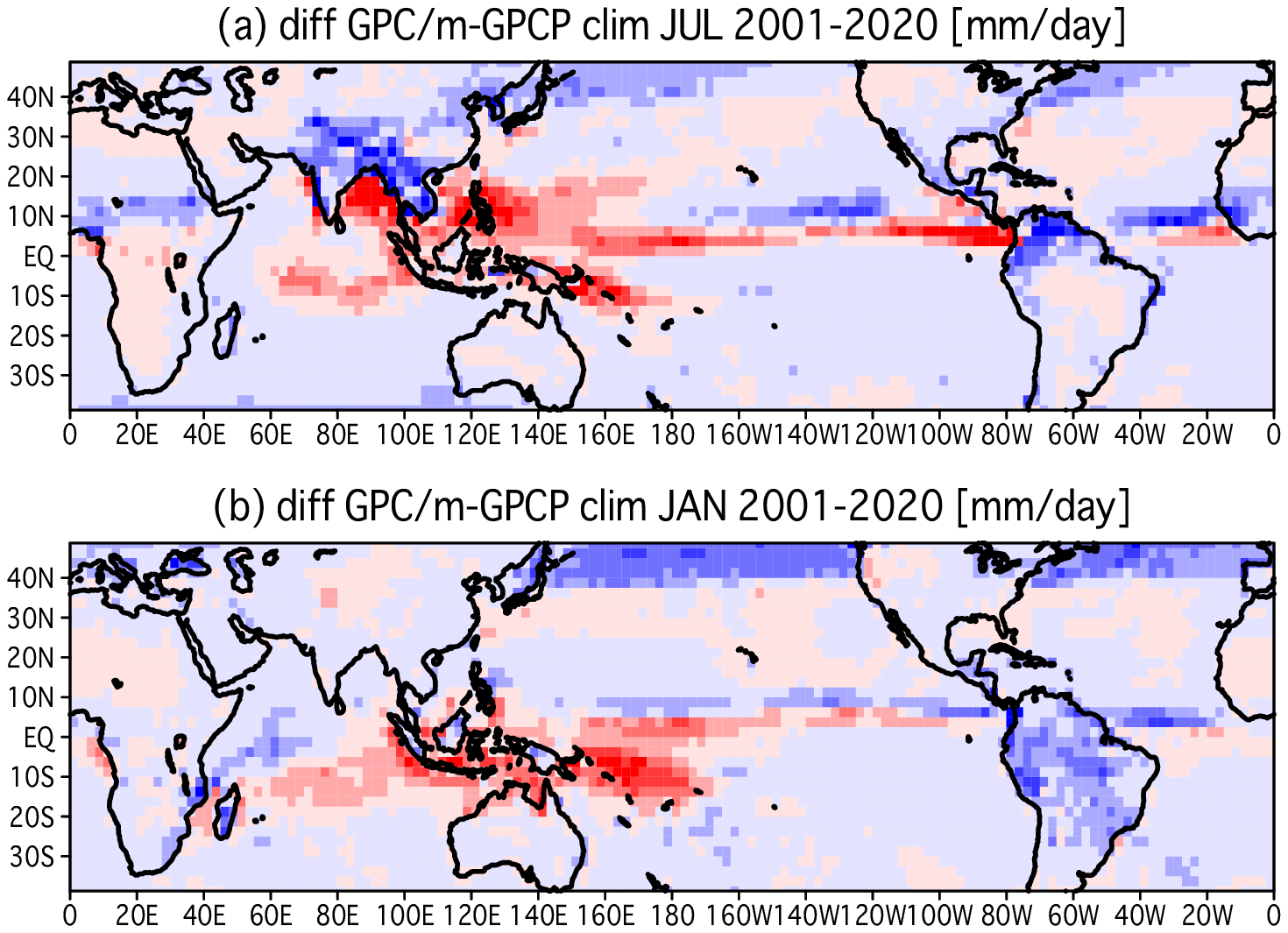}
      \noindent\includegraphics[width=22pc,angle=270]{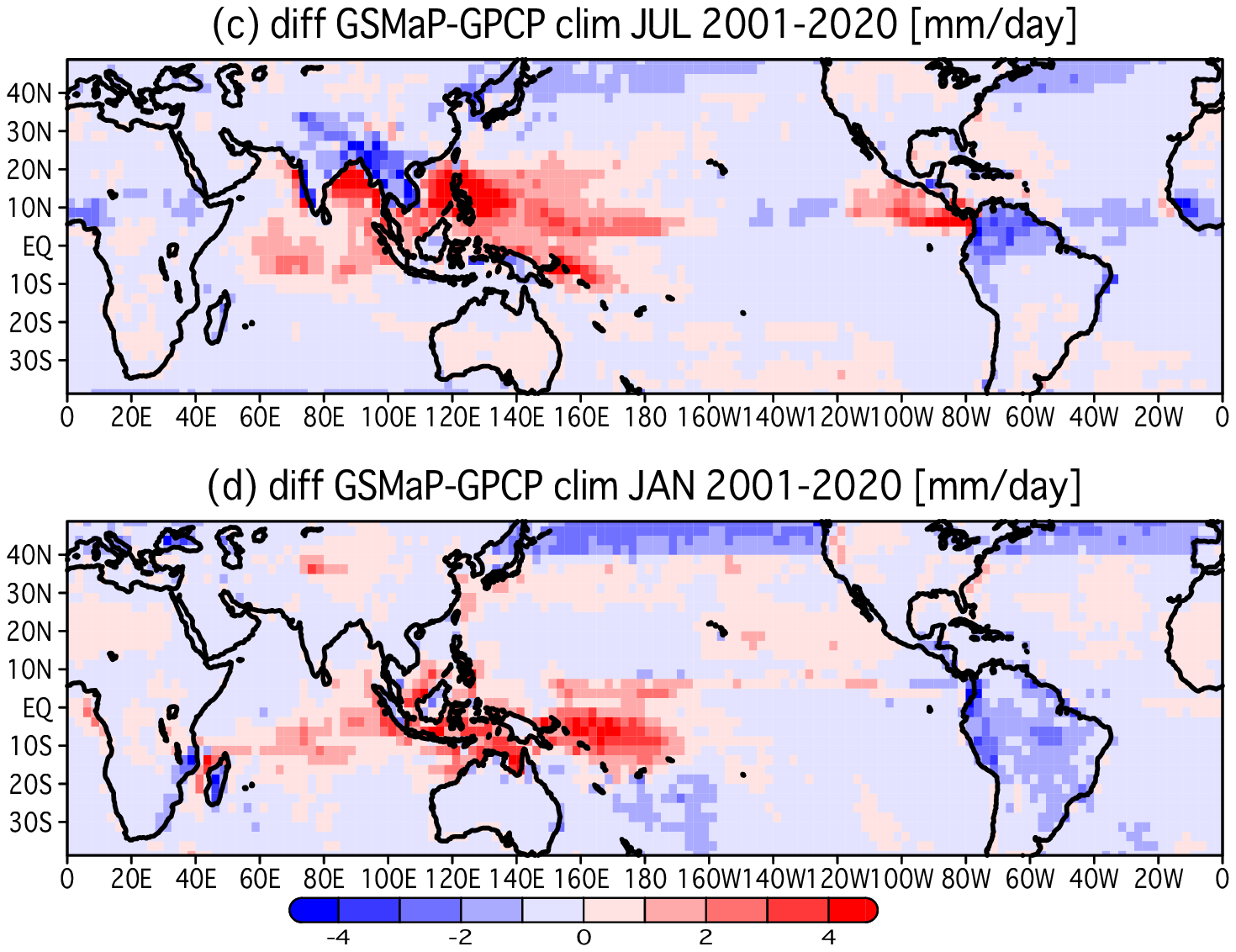}
      \caption{(a) Map of the climatological July precipitation difference between two datasets in GPC/m and GPCP from 2001 to 2020 . (b) Same as (a), but for January. (c) Same as (a), but for differences between GSMaP and GPCP for July. (d) Same as (c), but for January. Units are mm day$^{-1}$.}
\end{figure*}

Of course, there are some biases, but these biases are quite similar to the target precipitation values of GSMaP (Figure 2). A notable bias was observed over the South and Southeast Asian land regions, which can be considered as the differences of correction by surface precipitation observations between GSMaP and GPCP as shown in the differences between GPC/m and GPCP, and these between GSMaP and GPCP (Figure 2). Thus, our GPC/m was similar to GSMaP but not GPCP over these regions. Another bias over the mid-latitude ocean was also likely to be due to the treatment of sea ice between GSMaP and GPCP. Thus, the GPC/m values over north of 40$^{\circ}$N regions are currently uncertain.

These results have two implications that may be helpful in producing a dataset not limited to meteorology and climatology using these algorithms. First, the ensemble mean can reduce mean bias differences among individual predictions when applying an ensemble mean, such as a three-model ensemble. Second, the predicted values are well reproduced in the well-tuned statistical models; however, the predicted values are fairly fitted to the target values, including biased values.

The monthly precipitation of the 22-year climatology shows a systematic seasonal march, which is also similar to that of GPCP (Supplementary Video 1). Thus, the basic climatology of monthly and annual precipitation was well reproduced in GPC/m.

\subsection{Daily precipitation variability}  
It is pretty difficult to check the reproducibility of a daily sequence of precipitation as a figure; therefore, Supplementary Video 2 was added. Supplementary Video 2 shows two panels of observed and predicted precipitations. As an example, Supplementary Video 2 shows daily sequence of precipitations from January 1, 2020 to November 30, 2020.

Although it is not surprising that the observed precipitation was used as training data, the spatial patterns of precipitation were quite similar. For example, tropical cyclones over the northern Pacific Ocean sometimes develop over the tropical region of the western North Pacific, first moving westward, then northward, and then northeastward along the edge of the subtropical high over the western North Pacific. These tropical cyclones were reproduced and corresponded with the observations. However, the spatial pattern of the GPC/m precipitation is smoother than that of the GSMaP (also IMERG, not shown). The precipitation intensity of GPC/m is weaker than that of GSMaP, likely due to statistical models and ensemble mean.

Supplementary Video 3 shows the predicted precipitation for the period from January 1, 1979 to December 31, 1979, which is the period without GSMaP precipitation data. Interestingly, the behavior of tropical cyclones seems very similar to that of the observed precipitation. Because the basic precipitation system is considered to be determined climatologically, the reproduced precipitation variations cannot be unreasonable. Information from the reanalysis and OLR data likely generated daily precipitation variability.

\subsection{Long-term variability of the whole period of this dataset} 
This dataset is not recommended for the analysis of long-term changes and variability such as quasi-global mean precipitation. Figure 3 shows the 42-year time-series of the quasi-global mean precipitation zonally global from 40$^{\circ}$S to 40$^{\circ}$N. The time-series had a notable gap before and after 2000, which were the prediction and training periods, respectively. Therefore, the global mean total precipitation was not reproduced well by our statistical models. 

\begin{figure}
      \centering
      \noindent\includegraphics[width=15pc,angle=270]{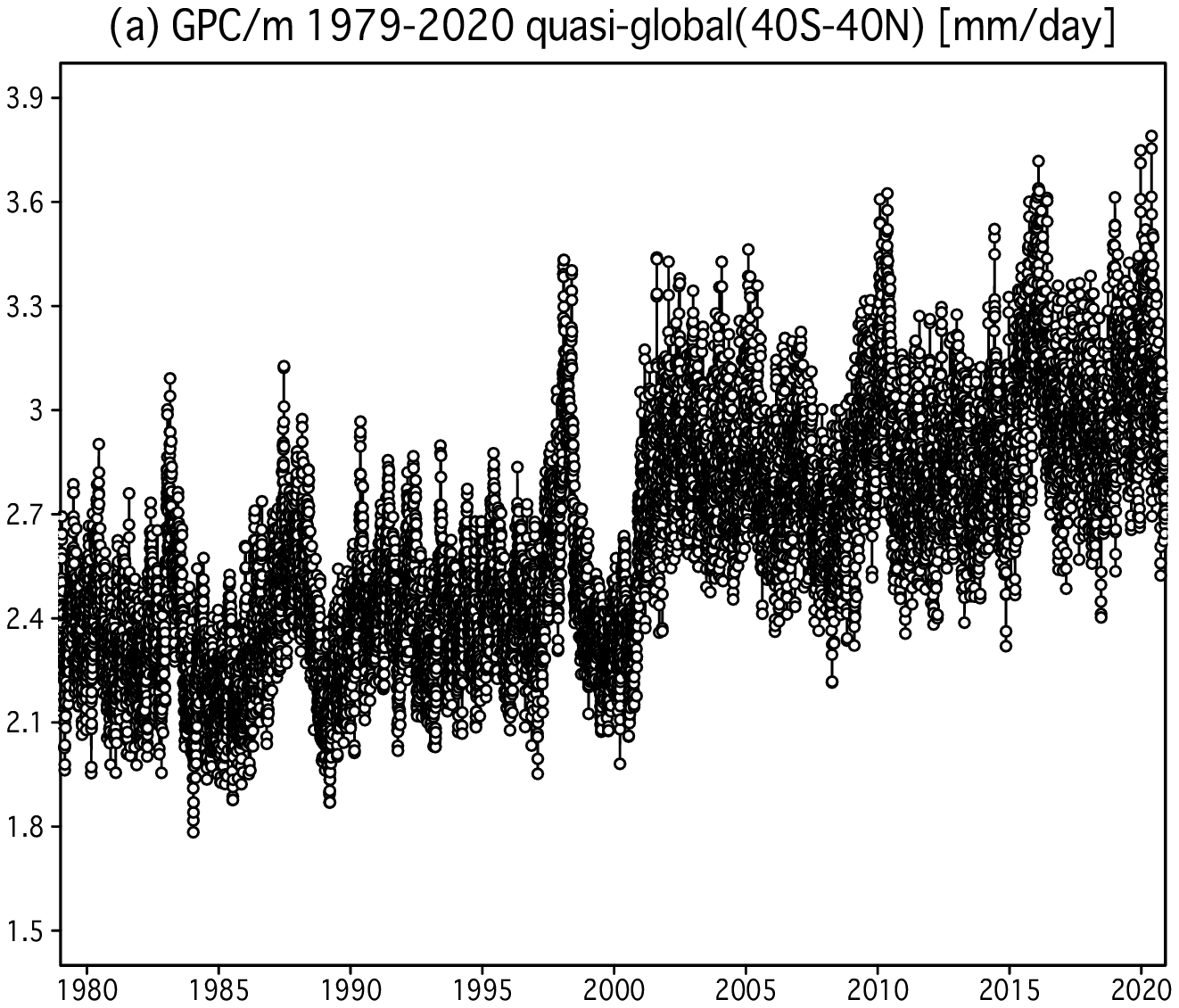}
      \noindent\includegraphics[width=15pc,angle=270]{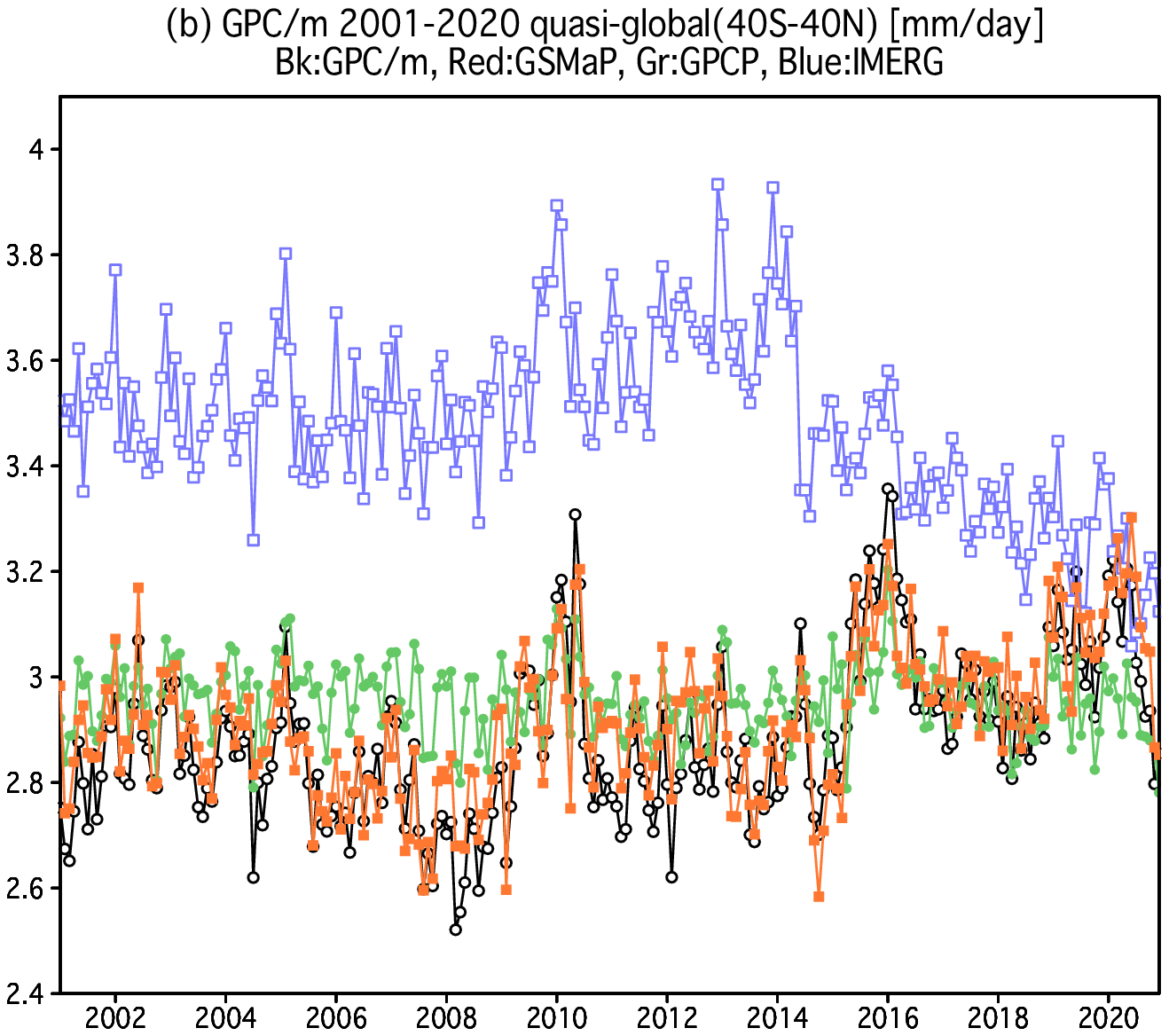}
      \caption{(a) Quasi-global mean precipitation of GPC/m. The mean precipitation was averaged zonally global from 40$^{\circ}$S to 40$^{\circ}$N. The training period is from 2001 to 2020, and the prediction period is from 1979 to 2000. (b) Quasi-global mean precipitation of GPC/m (black), GSMaP (red), GPCP(green), and IMERG (blue) over 20 years from 2001 to 2020. Units are mm day$^{-1}$.}
\end{figure}

However, the observed precipitation datasets, GSMaP, GPCP, IMERG, and our GPC/m, showed different long-term changes in precipitation variation (Figure 3b). The time-series of GSMaP and IMERG showed different long-term variations, while GPCP showed very small long-term variations. Our GPC/m dataset is quite similar to GSMaP because our GPC/m used GSMaP as reference data. These differences imply that the long-term trends in precipitation are still highly uncertain among the precipitation datasets when we examine the quasi-global mean (here, global in zonal direction and 40$^{\circ}$S to 40$^{\circ}$N in the north-south direction) value of precipitation. Understanding these differences and approaching the true values, which cannot be observed, are challenging tasks in climate research \citep[e.g.,][]{tak13,mas19}.

In addition, we examined the spatial patterns of long-term changes in GPC/m and GPCP. As indicated by the quasi-global mean variability (Figure 3a), positive signals were expected. The long-term changes in GPC/m showed an increase in precipitation over the tropical regions along the equator (Figure 4), which may be strongly positive biases, particularly over the Maritime Continent. Interestingly, long-term changes in GPCP also showed an increase in precipitation over the tropics, which is similar to GPC/m. However, the positive trends in precipitation were too strong for GPC/m. Although the quasi-global averaged values of precipitation showed different trends, the spatial patterns of long-term changes between GPC/m and GPCP were not inconsistent. This result implies that the quasi-global mean precipitation is associated with a positive bias in the long-term trend of GPC/m, which suggests that the regional patterns of precipitation trends in GPC/m can be mostly reproduced by subtracting the positive bias. It was noteworthy that our GPC/m precipitation was not nudged to another precipitation dataset during processing and post-processing.

\begin{figure*}
      \centering
      \noindent\includegraphics[width=28pc,angle=270]{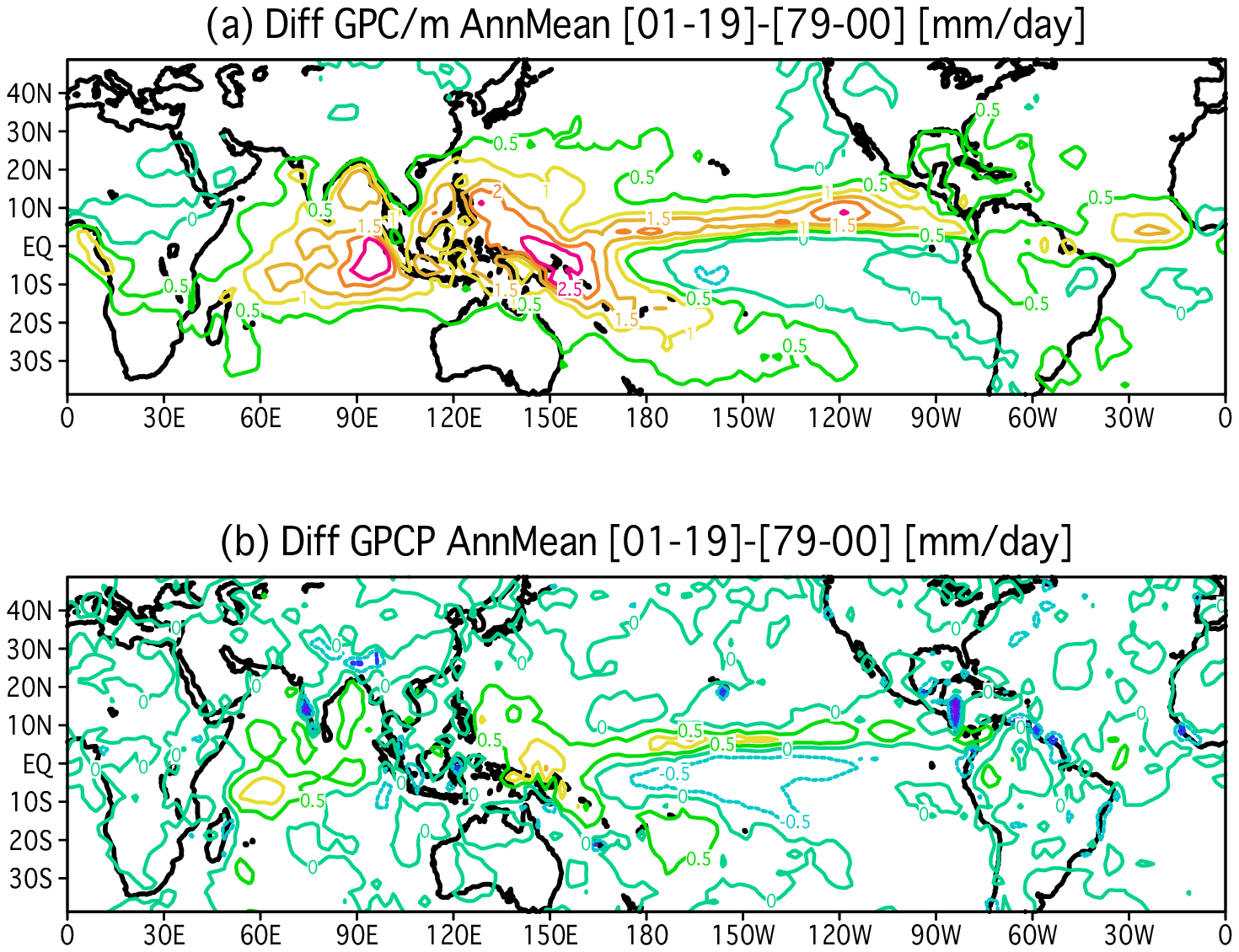}
      \caption{(a) Map of the annual precipitation difference of climatological means between two peirods from 2001 to 2020 and from 1979 to 2000 in GPC/m. (b) As for (a) but for GPCP. Units are mm day$^{-1}$.}
\end{figure*}

\subsection{Interannual variability}  

\subsubsection{ENSO influences}  

This section examines the interannual variability of precipitation related to El Ni$\widetilde{n}$o-Southern Oscillation (ENSO), although the interannual variations can be examined using monthly precipitation datasets such as GPCP. Here, northern fall and winter of 1979/1980, 1982/1983, 1986/1987, 1991/1992, and 1997/1998 were selected as the El Ni$\widetilde{n}$o years, whereas those of 1985/1985, 1988/1989, 1995/1996, 1998/1999 and 1999/2000 were selected as the La Ni$\widetilde{n}$a years (based on website of Japanese Meteorological Agency).

\begin{figure*}
      \centering
      \noindent\includegraphics[width=\textwidth,angle=0]{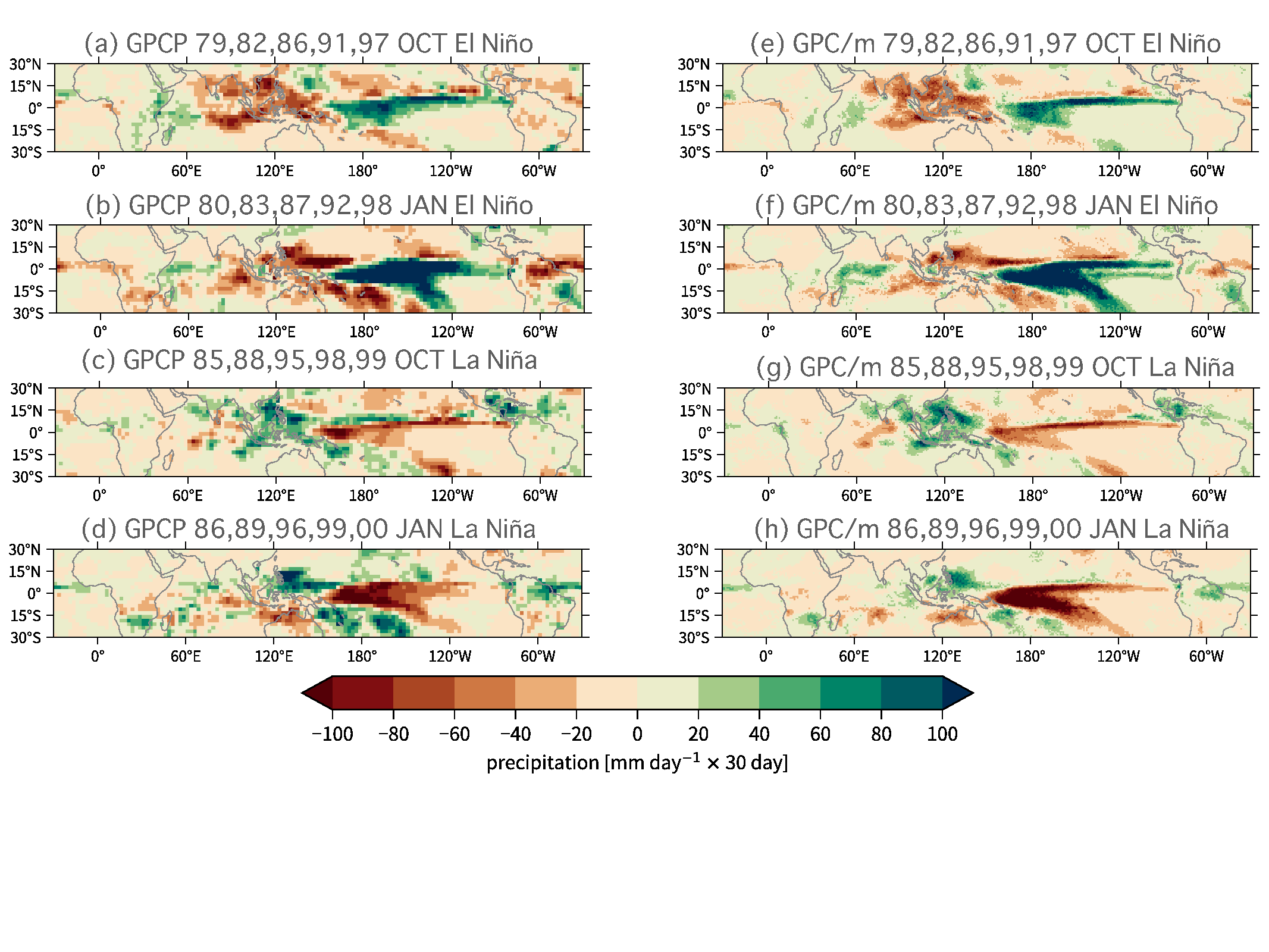}
      \caption{Composite maps of the monthly mean precipitation anomaly from climatological means for El Ni$\widetilde{n}$o years (a, b, e, f), and La Ni$\widetilde{n}$a years (c, d, g, h) during the 1979---2000 prediction period in GPCP (a, b, c, d) and GPC/m (e, f, g, h). October composites are (a, c, e, g) and January composites are (b, d, f, h). Units are mm 30-day$^{-1}$.}
\end{figure*}

The precipitation composite for El Niño years was very similar between the GPCP and GPC/m datasets (Figure 5). In northern fall and winter months, precipitation is lower over the Maritime Continent and higher over the Central Pacific. These spatial patterns were clearly predicted by GPC/m. In addition, the detailed structures of the precipitation anomalies were well predicted, indicating that the machine learning methods can generate the spatial precipitation patterns of El Niño. In addition, a spatial pattern of La Niña precipitation was also predicted (Figure 5).

\subsubsection{Decadal changes in precipitation over East Asia}  

\begin{figure}
      \centering
      \noindent\includegraphics[width=16pc,angle=270]{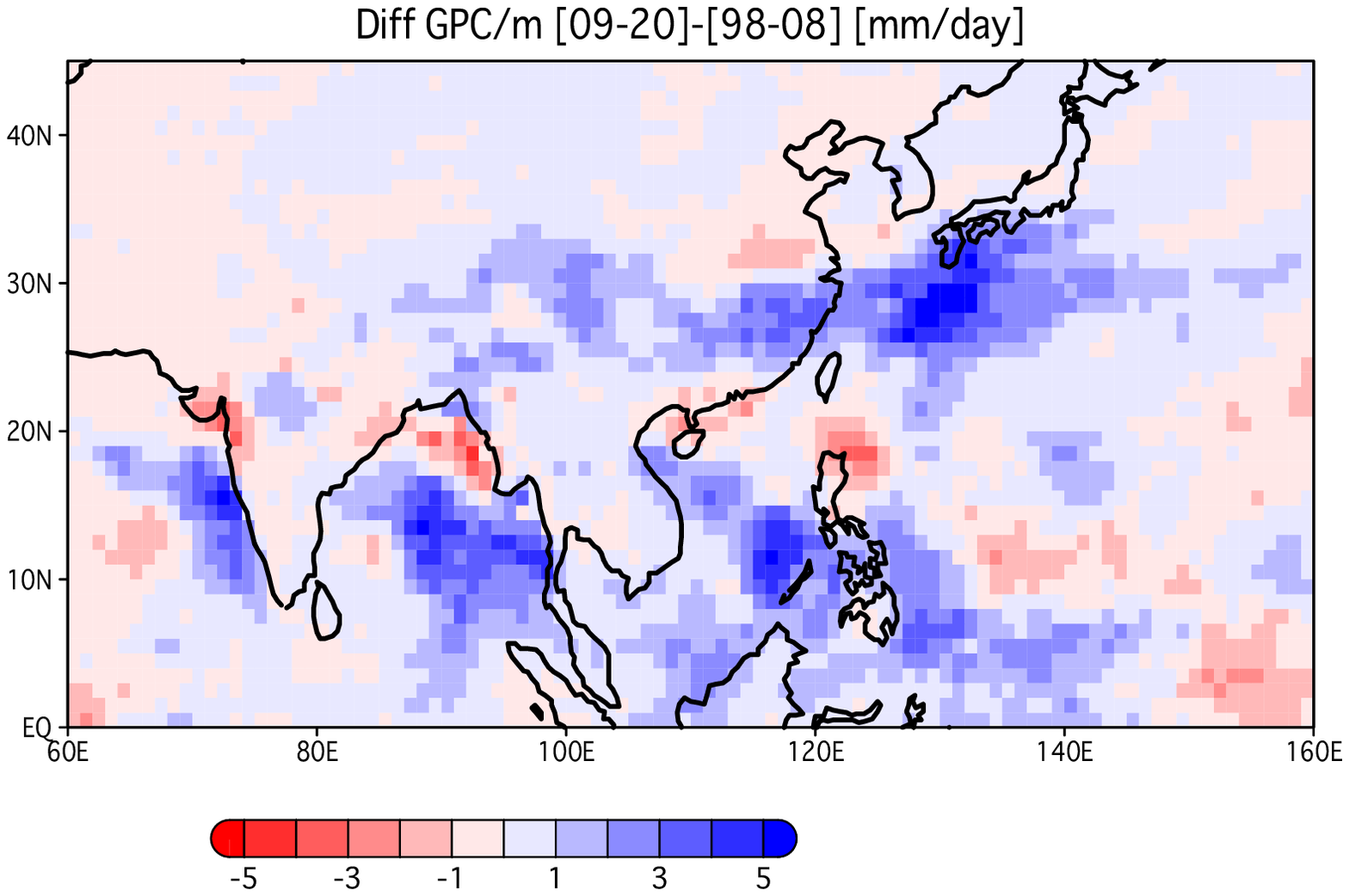}
      \caption{Map of GPC/m precipitation differences of the Meiyu-Baiu precipitation between 1998--2008 and 2009--2020 (see also Takahashi and Fujinami 2021). The Meiyu-Baiu period was defined as the period from the second half of June to the first half of July. Unit is mm day$^{-1}$.}
\end{figure}

For evaluation, decadal changes in precipitation over East Asia \citep[e.g.,][]{tak21} were reproduced using GPC/m. In the previous study, accurate precipitation estimations by TRMM-PR and GPM-DPR were used, which had also been partly used for GSMaP. Because the study focused on the dramatic fast seasonal changes of the Asian monsoon circulations and precipitation that the time-scale of the changes is shorter than a month, these daily data are very useful for analyzing non-calendar month-based precipitation data, such as synoptic and intraseasonal variability. Actually, most of the time-scales of meteorological and climatological phenomena are not followed by the month.

The results showed a notable increase in precipitation along the Meiyu-Baiu rainfall band from eastern China to southern Japan (Figure 6). The signal was almost similar to the observed precipitation signals by the TRMM-PR and GPM-DPR \citep{tak21}, indicating that GPC/m fairly reproduced the observed variations. However, because this period was mostly covered by the GSMaP precipitation, our precipitation dataset is more useful during the prediction periods. Nevertheless, additional evaluations are required to increase the reliability of analyzing 40-year trends and variability (e.g., Figure 4).

\subsection{Some significant precipitation variations}  

\subsubsection{Severe flood in Thailand 2011}  

For an evaluation from another perspective, we focused on the severe flooding event in 2011 in Thailand \citep[e.g.,][]{tak15}. These flooding events and high precipitation events were induced by intermittent active precipitation caused by several westward-propagating tropical cyclones \citep{tak08,tak09,tak20}. As shown in Supplementary Videos 1 and 2, the precipitation, which is likely to be associated with specific tropical cyclones, can be reproduced in GPC/m. 

\begin{figure}
      \centering
      \noindent\includegraphics[width=16pc,angle=270]{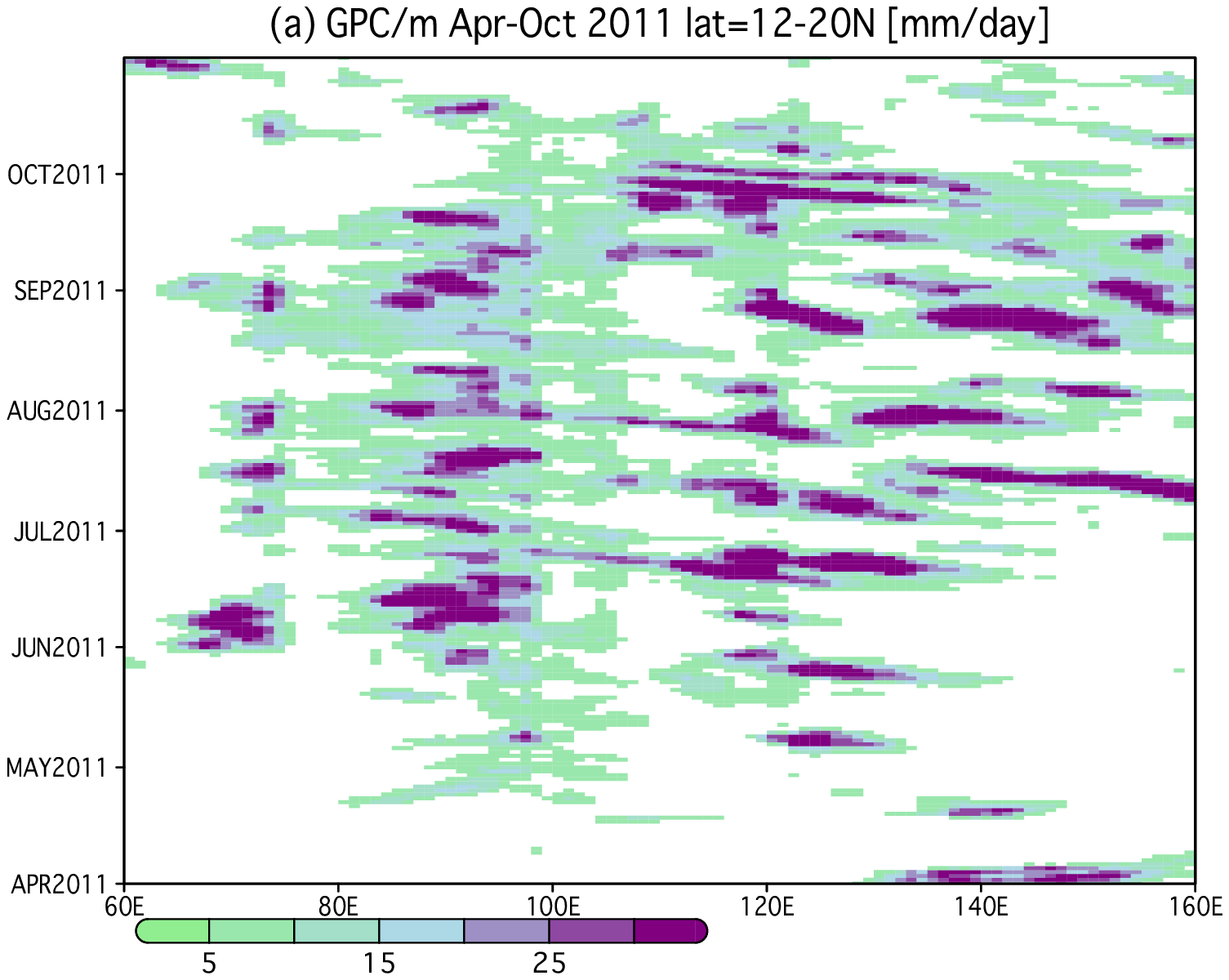}
      \noindent\includegraphics[width=16pc,angle=270]{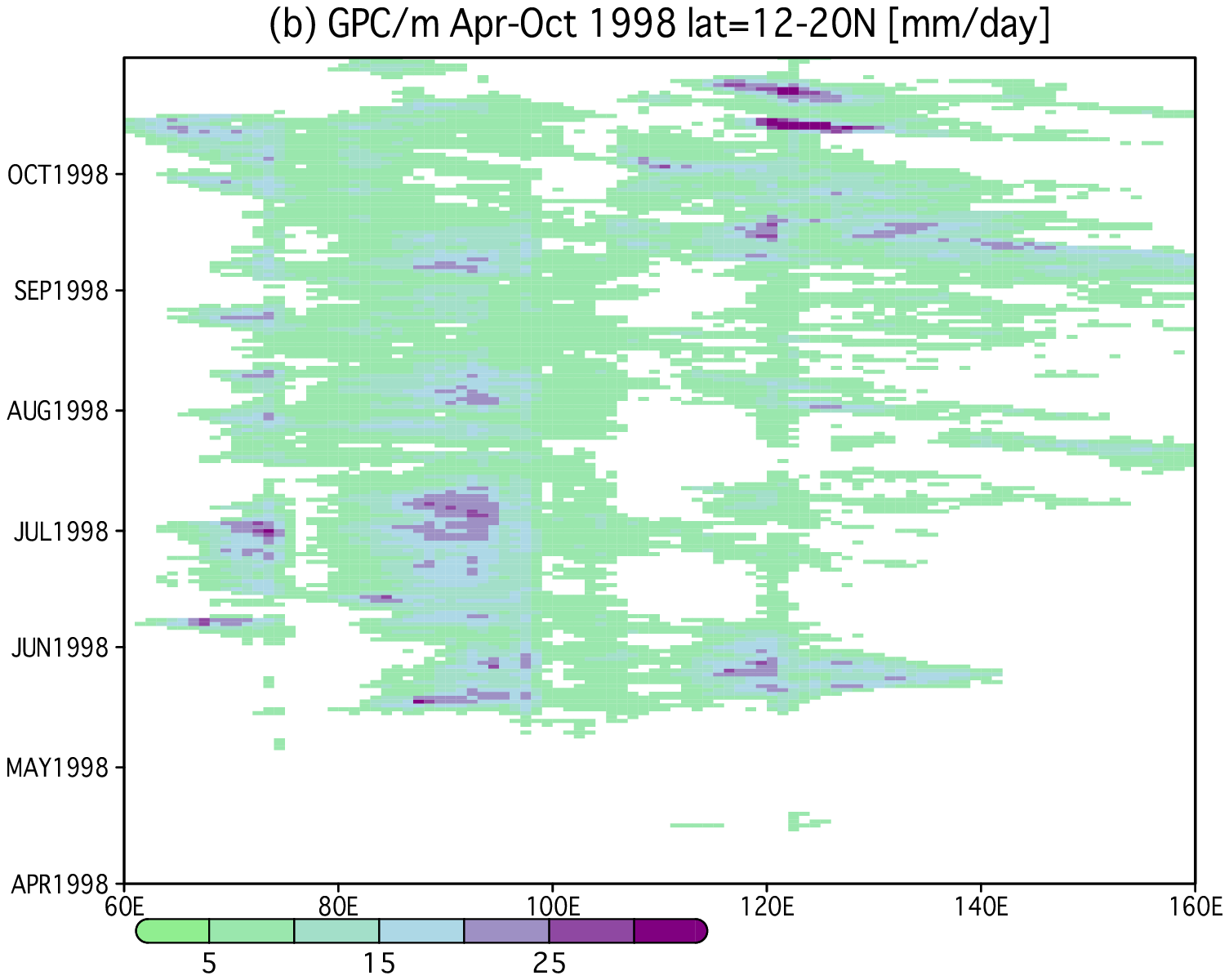}
      \caption{Time-longitude cross section generated from the GPC/m daily precipitation in (a) 2011 and (b) 1998 along the latitudinal band between 12$^{\circ}$N and 20$^{\circ}$N. Units are mm day$^{-1}$.}
\end{figure}

Westward-propagating precipitation systems along the latitudinal band between 12$^{\circ}$N and 20$^{\circ}$N were frequently observed in GPC/m (Figure 7). These precipitation systems corresponded well with the results of a previous study \citep{tak15}. As a contrasting example of a dry year in Thailand, 1998 showed long precipitation breaks, indicating that intraseasonal variations in precipitation were reproduced. Thus, the interannual precipitation variations associated with these precipitation events and precipitation due to tropical cyclones were reproduced well. However, as 2011 was included in the training period, it may be necessary to evaluate another case during the prediction period. Also, the underestimation in precipitation in GPC/m during the prediction period overemphasized less precipitation in 1998. At least, in Supplementary Video 2, some tropical cyclones, including weaker tropical disturbances (not typhoons), traveling westward, which are originated from the South China Sea and western North Pacific, to the Indochina Peninsula are observed after the monsoon onset (approximately June and July 2020).

\subsubsection{Severe flood in Pakistan 2022}   

\begin{figure*}
      \centering
      \noindent\includegraphics[width=26pc,angle=270]{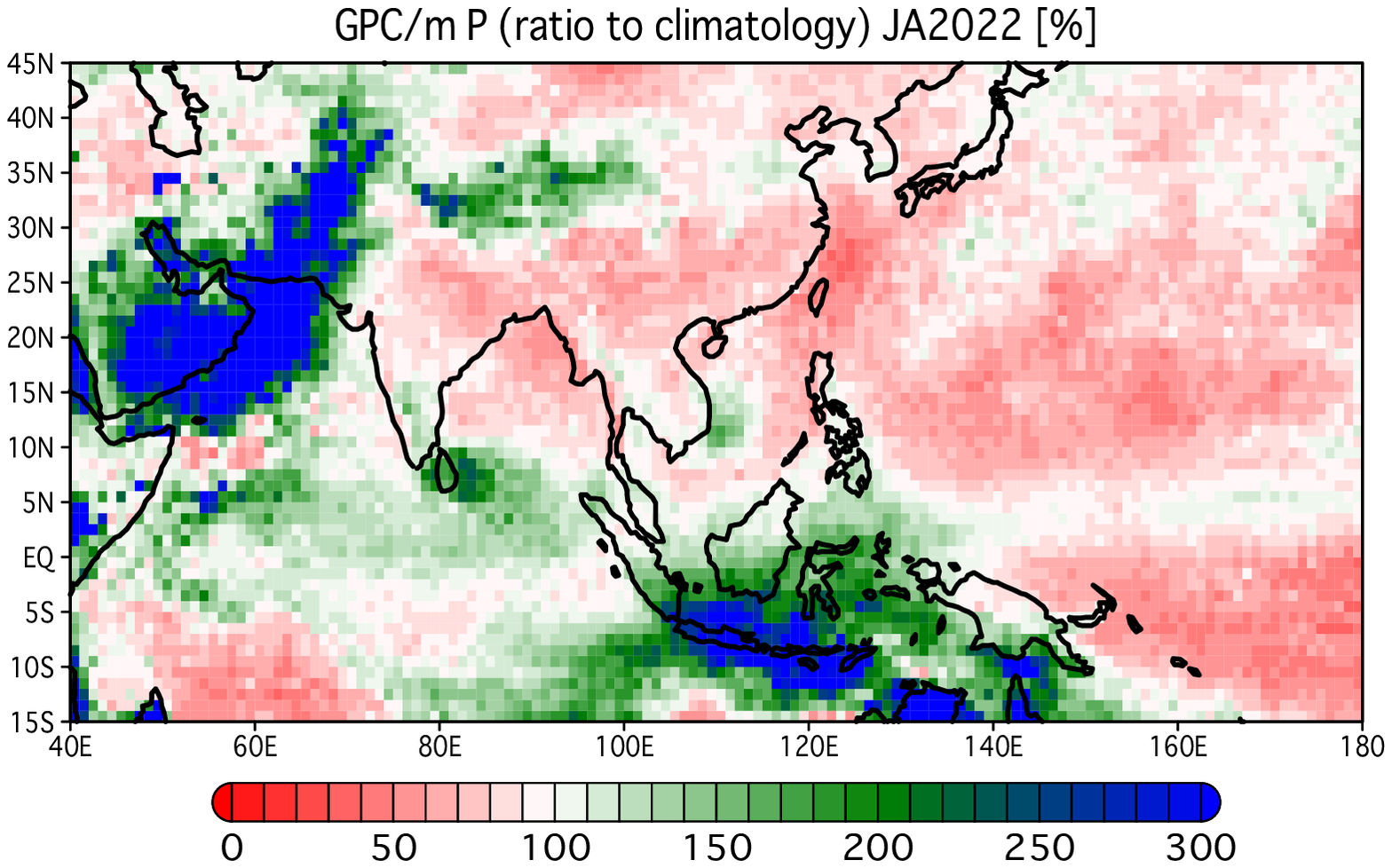}
      \caption{Spatial map of precipitation ratio in July and August 2022 to the climatology in percentage. Climatology was calculated from 1979 to 2022. 100\% showing in a year indicates that the year is a normal precipitation year.}
\end{figure*}

As a severe flooding case of the non-training period, the 2022 Pakistan flood \citep{tak24} was examined, which were due to several active tropical cyclones staying for a long time along the South Asian monsoon trough. The active monsoon trough conditions can be understood as the long active phase of the Indian monsoon. As this period was not included in our dataset, we produced precipitation data from 2021 to 2022 using statistical models trained by the data from 2001 to 2020, which did not include 2022. 

The spatial map of precipitation showed a significant precipitation anomaly, which was reproduced, even though Pakistan has never experienced such severe precipitation events (Figure 8). The precipitation in July and August reached to 300\% to the climatological values, which is consisitent with the observational analysis \citep{tak24}. Thus, although the training period does not have severe events similar to 2022, the statistical model has a potential to reproduce the severe precipitation. However, total values can possibly be reproduced, but it is not easy to reproduce daily precipitation of each day, although it can be reproduced when precipitation events tend to occur or not. This result implies that machine learning methods can reproduce part of future precipitation events from similar present precipitation events from atmospheric circulations data. Nevertheless, because OLR observation data can be significant for precipitation predictions, it may not be easy to predict future precipitation because of the absence of future OLR observation.

\section{Discussion and dataset features}  
\subsection{Characteristics of machine-learning-based precipitation}  

As shown in the composite analysis of GPC/m based on ENSO, these statistical analyses show great potential for reproducibility. In general, our GPC/m precipitation datasets are useful for statistical analyses of synoptic, intraseasonal, seasonal, and interannual precipitation variations that we meteorological and climatological scientists most commonly study. However, it was difficult to reproduce specific precipitation events. For example, a specific heavy precipitation event during the 2022 Pakistan floods might not be predicted, although two months of active precipitation activity associated with tropical disturbances were predicted

As mentioned in Section 3.1, if the targets have biases, the predictions will also inherit the biases. However, the predictions made by each machine learning method have different individual biases that can be reduced. However, if each machine learning statistical model is perfect, the prediction converges to the target values.

Also, the ensemble method can reduce the random errors. As I will mention below, a specific precipitation event have less reproducibility than that of statistics, which is partly improved by the ensemble method. Thus, to reduce a bias of a target value, it should be considered in the future, such that multiple kinds of target values are used. Nevertheless, it should be carefully considered what is the possible realistic value.

Also, the ensemble method can reduce the random errors. As I will mentioned below, a specific precipitation event has less reproducibility than statistics, which can be partially improved by the ensemble method. Thus, to reduce a bias of a target value, it should be considered in the future, such that multiple kinds of target values are used. Nevertheless, we should carefully consider what the realistic values are.

\subsection{Unsutability for the analysis of specific precipitation events}  

This section describes the difficulties of the analysis using this dataset. In short, because thesethis data werewas produced by the statistical models, the model- predicted values showedare statistically better performance. Specifically, as mentioned above, thesethis data haves a great potential for the statistical analysis, such as composite and correlation analyseis, but there are difficulties in predicting a specific precipitation event, especially for a short-period event or a local-scale precipitation events. This is not only due tofrom the insufficient high- resolution capability, but also because the prediction of each precipitation event is generally not better than statistics, such as an average. In other words, the deterministic prediction performance wais low.

Another concern is climatological small-scale or sub-daily precipitation (e.g., the diurnal precipitation cycle). For example, higher precipitation along the western coast of the Indochina Peninsula during the summer monsoon season \citep[e.g.,][]{tak10a,tak10b}, along the southwestern coast of Sumatra Island during the northern fall to winter season \citep[e.g.,][]{hir05}, were well reproduced.

However, these precipitation amounts are induced by diurnal precipitation cycles, which must not be depicted in statistical models because all models predict only one-day mean precipitation. These significant precipitation amounts have been reproduced by statistical models; however, it should be noted that statistical models do not consider the physical processes of precipitation and simply predict precipitation amounts statistically.

\section{Summary and Future}  
This study produced a daily precipitation dataset for approximately 45 years, from 1979 to 2022, using three different machine learning methods. The training period for the development of the statistical model was 20 years, from 2001 to 2020. The spatial resolution was moderate, which can be useful for climatological analysis at intraseasonal and interannual time-scales. The input datasets were OLR and atmospheric reanalysis data on a daily basis. Using RF, XGB, and CNN, we predicted and averaged the three predicted precipitation values. Daily precipitation values were predicted for individual grids using individual statistical models and their sets of parameters. Therefore, individual grids have different statistical relationships between precipitation and input.

The predicted products were evaluated by a comparison with the observational datasets. Precipitation variations at synoptic, intra-seasonal, seasonal, and interannual time-scales, reported in previous studies, were evaluated. We then concluded that our GPC/m is useful for climatological statistical analyses of precipitation variations at these various time-scales. However, it can be difficult to analyze specific localized severe weather events using this dataset. Although this does not mean that the predicted values are apart from the observations, they are simply calculated statistically. In other words, deterministic prediction skills are insufficient because these data were produced using statistical approaches. Therefore, it is sometimes difficult to use these data to analyze specific heavy precipitation events. In this manner, this study also discussed the difficulties associated with GPC/m. These difficulties are common in non-physical statistical model datasets.

This study also discussed the difficulties associated with GPC/m, which may be common in non-physical statistical model datasets. This dataset included other points of insufficient quality. For example, the long-term trend of the quasi-global mean precipitation differs from that of GPCP. However, other datasets exhibited different trends. In addition, there was a significant gap before and after December 2000 due to the boundary between the prediction and training periods. Finally, this dataset may contain unexpected problems, biases, and errors. Thus, the author hopes that users will report their problems to him.

The purpose of this dataset is a challenge to produce a climatological dataset by reducing artificial biases as much as possible for discussions on climatology, climate variability, and climate changes. Also, I hope that this dataset may contribute to improving current precipitation datasets, which are based on physical or researcher-explaining algorithms.  

The current GPC/m version is 1-2024 (Takahashi, 2024b). This will be updated in the future with added value.

\section{Data distribution} 
We have distributed our GPC/m dataset via Zenodo, an open-access repository for sharing and preserving research outputs \citep{tak24b}. To facilitate the analysis of the dataset, it is distributed in the Network Common Data Form (netCDF) format and the Grid Analysis and Display System (GrADS) format (with a control file). When GPC/m is updated, new data will be available from Zenode \citep{tak24b}.

\section*{Acknowledgement}
Climate Prediction Center (CPC) Daily Blended Outgoing Longwave Radiation (OLR) - 2.5-degree data provided by the National Oceanic and Atmospheric Administration (NOAA) Physical Science Laboratory (PSL), Boulder, Colorado, USA, from their website at \url{https://psl.noaa.gov}. The Global Satellite Mapping of Precipitation (GSMaP), Global Precipitation Climatology Project (GPCP), and Japanese Reanalysis 55 (JRA55) data were downloaded from the Japan Aerospace Exploration Agency (JAXA), the National Aeronautics and Space Administration (NASA), and the Research Data Archive (RDA) at the National Center for Atmospheric Research (NCAR), respectively. IMERG was also downloaded from NASA. For the random forest (RF), XGBoost (XGB), and convolutional neural network (CNN) programs, we used the Scikit-learn, xgboost, and TensorFlow packages, respectively. For parallel script programming, we used the GNU Parallel by Dr. Ole Tange. To explore the functions of each package related to machine learning and parallel programming, we partially used ChatGPT prompts while verifying them with other resources and tests. We thank Sho Kitabayashi for helpful technical supports. This work was partly supported by the Japan Society for the Promotion of Science KAKENHI grants 21K18403, 22H00037, and 24K21389, as well as the 3rd Earth Observation Research Announcement (EORA-3) ER3GPF012 from the Japan Aerospace Exploration Agency.



\section*{Supplementary Material}

\subsection*{Supplementary Video}

\noindent
The following supplementary videos provide additional visual explanations of the results discussed in this paper.

\begin{itemize}
    \item \textbf{Supplementary Video 1:} Visualization of the climatological seasonal march in precipitation of GPC/m (upper panel) and GPCP (lower panel). The video shows the climatological precipitation in order from January to December. Unit is mm day$^{-1}$.
    \item \textbf{Supplementary Video 2:} Visualization of daily precipitation of GPC/m and GSMaP from January 1, 2020 to November 30, 2020, which is during the training period. Unit is mm day$^{-1}$.
    \item \textbf{Supplementary Video 3:} Visualization of daily precipitation of GPC/m from January 1, 1979 to December 31, 1979. Unit is mm day$^{-1}$.
\end{itemize}

\bibliography{910gpcm}

\end{document}